# Altermagnetic Skyrmions in 2D Lattices Exhibiting Anisotropic Skyrmion Hall Effect

Kaiying Dou, Zhonglin He, Wenhui Du, Ying Dai*, Baibiao Huang and Yandong Ma*

School of Physics, State Key Laboratory of Crystal Materials, Shandong University, Shandanan Street 27, Jinan 250100, China

*Corresponding author: daiy60@sina.com (Y.D.); yandong.ma@sdu.edu.cn (Y.M.)

**Abstract**

Anisotropic skyrmion Hall effect (A-SkHE) in two-dimensional (2D) magnetic systems represents a captivating phenomenon in condensed-matter physics and materials science. While conventional antiferromagnetic systems inherently suppress this effect through parity-time symmetry-mediated cancellation of Magnus forces acting on skyrmions, A-SkHE is primarily confined to ferromagnetic platforms. Here, we present a paradigm-shifting demonstration of this phenomenon in spin-splitting 2D antiferromagnets through the investigation of altermagnetic skyrmions. Combining comprehensive symmetry analysis with theoretical modeling, we elucidate the mechanism governing A-SkHE realization in 2D altermagnetic systems and establish a quantitative relationship between the transverse velocity of altermagnetic skyrmions and applied current orientation. Using first-principles calculations and micromagnetic simulations, this mechanism is further illustrated in a prototypical altermagnetic monolayer $V_2SeTeO$. Crucially, we identify that the $[C_2 \| C_{4z}\mathbf{t}]$ symmetry-protected anisotropic field serves as the critical stabilizer for maintaining the A-SkHE in this system. Our results greatly enrich the research on 2D altermagnetism and skyrmions.

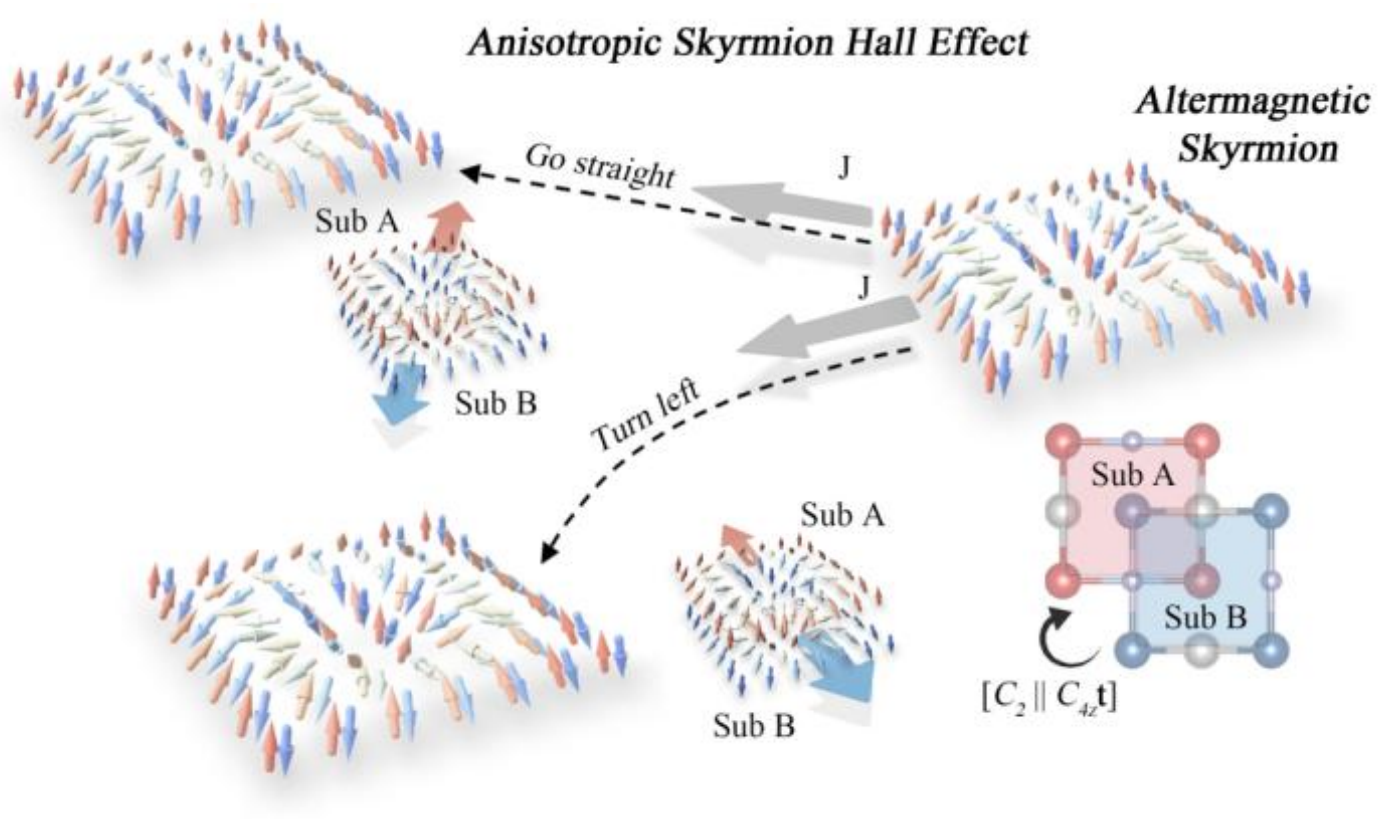

*Table of Content*

## Introduction

Topological spin textures, characterized by their chiral quasiparticle nature and nontrivial real-space topology, have emerged as a frontier research field due to both fundamental scientific significance and advanced spintronic applications [1-5]. A hallmark phenomenon in these systems is the skyrmion Hall effect (SkHE), which manifests the travel of these quasiparticles with curved trajectories away from the direction of an applied electric current [6-9]. This emergent transport phenomenon has been predominantly observed in ferromagnetic (FM) systems across various topological spin configurations [6-8]. Particularly intriguing is the SkHE in elliptical skyrmions and antiskyrmions, wherein the skyrmion Hall angle evolves dynamically with the current orientation rather than remaining constant [10-15], giving rise to the concept of anisotropic SkHE (A-SkHE). Undoubtedly, A-SkHE opens new dimensions for exploring the physics of chiral quasiparticles and offers unprecedent opportunities for advancing topological spintronics [16-19].

Unlike FM counterparts, antiferromagnetic (AFM) systems without net magnetization and associated stray fields have been demonstrated to be able to address critical challenges in device scaling, enabling enhanced integration density and improved operational robustness against external perturbations [20-25]. Furthermore, their inherent ultrafast spin dynamics operating in the terahertz frequency regime also enables fast operations [26-28]. Considering these compelling characteristics, the realization of A-SkHE in AFM platforms is highly expected to advance A-SkHE-based devices [29-31]. Nevertheless, due to the parity-time (PT) symmetry-enforced cancellation of Magnus forces acting on skyrmions, conventional AFM architectures fundamentally preclude both SkHE and its anisotropic variant [32,33]. Currently, A-SkHE is predominantly confined to FM systems, and how to expand A-SkHE to AFM coupled systems remains a persistent grand challenge in condensed matter physics [34].

Recently, a novel magnetic category termed altermagnetism, uniquely unifying spin-splitting bands with compensated AFM ordering, has been proposed theoretically [35-37] and verified experimentally [38-40]. In this work, we report the identification of A-SkHE in spin-splitting two-dimensional (2D) antiferromagnets through the discovery of stabilized altermagnetic (ATM) skyrmion phases. By integrating symmetry analysis with theoretical modeling, we decode the physics for the emergence of A-SkHE in 2D ATM lattices and unveil a precise quantitative relationship between the transverse velocity of ATM skyrmions and the orientation of applied current orientation. Based on first-principles calculations and micromagnetic simulations, we further demonstrate the validity of this mechanism in monolayer $V_2SeTeO$. Our dynamical analysis reveals that the skyrmion transverse velocity exhibits strong directional dependence on current orientation, with the $[C_2 \| C_{4z}\mathbf{t}]$ symmetry-protected anisotropic field emerging as the essential stabilizing factor for sustaining A-SkHE. These findings provide critical insights coupling A-SkHE with AFM characteristics and would advance 2D topological spintronics.

**Results and Discussion**

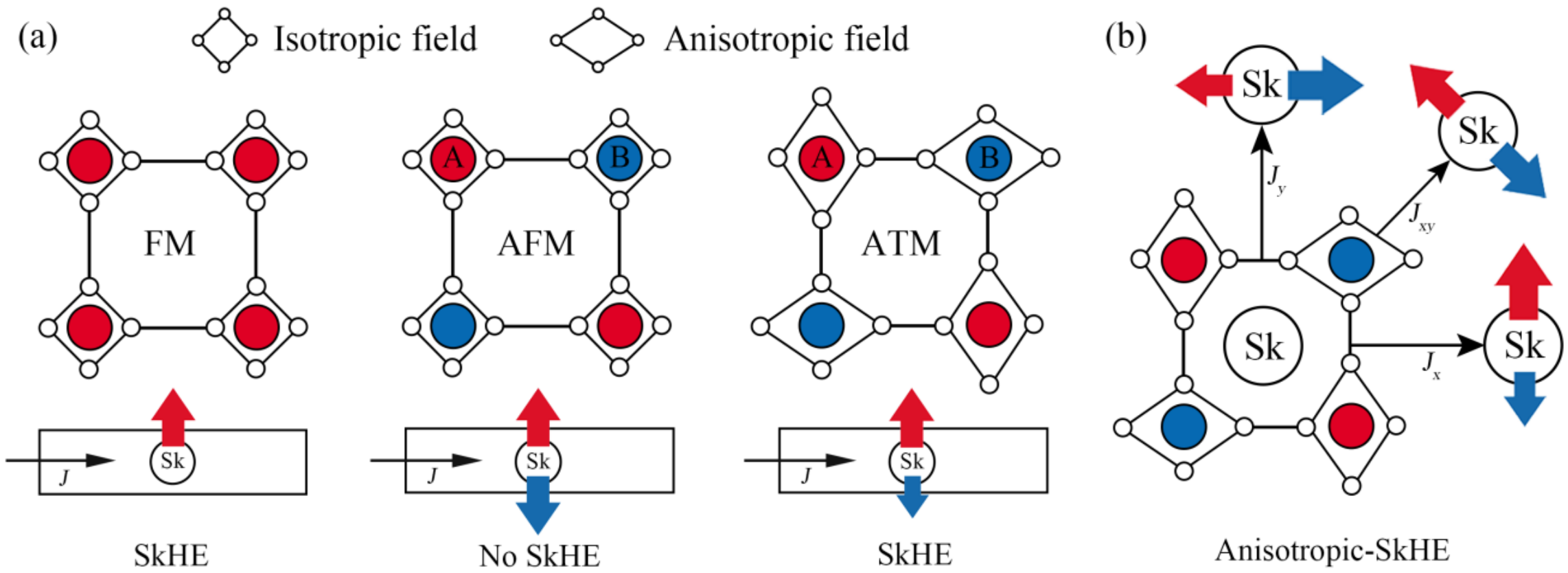

**Fig. 1** (a) Schematic diagrams of different magnetic configurations and their corresponding SkHE. (b) Schematic diagrams of A-SkHE in 2D altermagnets. Red arrows represent the transverse velocity of sublattice A, while blue arrows indicate the transverse velocity of sublattice B.

To set the stage, we first conduct a generalized model to investigate the current-induced dynamics of skyrmions based on standard Landau-Lifshitz-Gilbert (LLG) equation [41-43]. With disregarding the annihilation or shape deformation of topological spin textures, LLG equation combined with spin-transfer torque (STT) can be reduced to Thiele equation as follows [44-46]:

$$\mathbf{G} \times (\boldsymbol{u}-\boldsymbol{v}) + \overleftrightarrow{D} \cdot (\beta \boldsymbol{u} - \alpha \boldsymbol{v}) + \nabla V = 0 \tag{1}$$

Here, the three terms describe the Magnus force, dissipative force and confining force, respectively. The Magnus force is directly associated with the transverse displacement of skyrmions. $\mathbf{G} = \begin{pmatrix} 0 \\ 0 \\ g \end{pmatrix}$ is the gyromagnetic coupling vector, wherein $g$ = -4π$Q$ and $Q$ is topological number described as $Q = \frac{1}{4\pi}\int \boldsymbol{m} \cdot \left(\frac{\partial \boldsymbol{m}}{\partial x} \times \frac{\partial \boldsymbol{m}}{\partial y}\right) \mathrm{d}x\mathrm{d}y$. $\boldsymbol{v} = \dot{\mathbf{R}}$ is the observable velocity of skyrmions and $\mathbf{R}$ represents the location of skyrmions. $\boldsymbol{u}$ is the velocity of spin-polarized current. When spin-polarized current flows through skyrmion, the spin angular momentum of electron is transferred to skyrmion, generating a STT, which drives its motion. The magnitude of $\boldsymbol{u}$ is directly proportional to the applied current density $j_s$ and its direction is collinear with the electron flow. The dissipative force represents the effect of drag force on skyrmion motion. $\overleftrightarrow{D} = 4\pi\begin{pmatrix} D_{xx} & D_{xy} \\ D_{yx} & D_{yy} \end{pmatrix}$ is dissipative force tensor, wherein $D_{ij} = \frac{1}{4\pi}\int \partial_i \boldsymbol{m}\cdot\partial_j \boldsymbol{m}\mathrm{d}x\mathrm{d}y\,(i, j = x \text{ or } y)$. $\nabla V$ is the force arises from confining potential, which can be induced by sample edges, defects or other external potential landscapes. $\alpha$ is the Gilbert damping constant and $\beta$ is the strength of the non-adiabatic torque.

In conventional isotropic FM systems, spins adopt parallel alignment with equivalent coordination environments for each magnetic atom [35,36]. This configuration creates an isotropic field that acts uniformly on all magnetic sites, as schematically depicted in the left panel of **Fig. 1a**. Then we set $D_{xx} = D_{yy} = D$, $D_{xy} = D_{yx} = 0$ and $V = 0$ [45,46]. When an electric current flows through these systems, the skyrmion velocity is derived as $\boldsymbol{v} = \begin{pmatrix} v_x \\ v_y \end{pmatrix} = \frac{u}{\alpha^2 D^2 + g^2}\begin{pmatrix} \alpha\beta D^2 + g^2 \\ (\alpha\text{-}\beta)\, gD \end{pmatrix}$, where $x$-axis is defined along the direction of current flow and $y$-axis is oriented perpendicular to it [45]. Obviously, the skyrmion velocity is independent of the current direction, suggesting the isotropic SkHE.

In AFM systems, the magnetic structure can be formally described by two FM sublattices coupled via a combination of time-reversal ($\boldsymbol{T}$) symmetry and spatial operations such as inversion ($\boldsymbol{P}$) or translation (**t**) symmetries. These composite symmetries are conventionally denoted as $[C_2 \| \mathbf{t}]$ and $[C_2 \| \boldsymbol{P}]$, where operators to the left of the double bar operate exclusively in spin space, while those to the right act solely in real space. When an AFM skyrmion is generated, its constituent sublattices host FM skyrmions carrying opposite topological charges ($Q_A = -Q_B$ with A and B labeling the two sublattices) and STT is applied to two coupled FM sublattices (A and B) individually. The effective force on AFM skyrmion arises from the interplay of torques on these coupled sublattices. The validity of this method has been demonstrated in previous work [33] and widely adopted for the following simulation on AFM systems [45,47,48].

This AFM topological configuration ensures synchronized longitudinal motion ($v_x^A = v_x^B$) but antiparallel transverse velocities ($v_y^A = -v_y^B$) under current-driven conditions. The resultant transverse velocities of the two sublattices exactly compensate each other, leading to complete cancellation of net transverse motion for the composite AFM skyrmion. This mutual compensation mechanism effectively suppresses the SkHE, as illustrated in the middle panel of **Fig. 1a**. Although AFM textures theoretically possess inertial effect [49], the inertial term can be neglected here because it contributes to the dynamics only on short time scales [33]. Meanwhile, the study of skyrmion Hall effect mainly focuses on the steady-state dynamics, and it is consistent with recent works on AFM and ATM skyrmions [33,34].

The preceding analysis implies that inducing the SkHE in AFM coupled systems requires establishing nonreciprocal transverse velocity components between the two sublattices. This necessitates symmetry breaking of the original [$C_2$‖**t**] or [$C_2$‖P] constraints governing sublattice coupling. Crucially, while lifting these symmetries, the system must maintain zero net magnetization to prevent stray field generation, mandating compensating symmetry operations to preserve global AFM order. Our symmetry analysis demonstrates that these dual requirements can be simultaneously satisfied by (i) substituting isotropic magnetic interactions with anisotropic fields and (ii) establishing discrete rotational symmetry ($C_n$**t**) that couples sublattices while breaking both ***P*** and **t** symmetries. This theoretical framework identifies 2D ATM lattice as ideal candidates fulfilling these engineered symmetry conditions.

As a representative example, we consider the skyrmion realized in 2D ATM lattice with sublattices interconnected by [$C_2$‖$C_{4z}$**t**] symmetry, as illustrated in the right panel of **Fig. 1a**. Owing to the [$C_2$‖$C_{4z}$**t**] symmetry, the topological spin configuration of each sublattice can be effectively regarded as an elliptical skyrmion with intrinsic anisotropy. In this sense, we obtain $D_{xx} \neq D_{yy}$, $D_{xy} \approx D_{yx} \approx 0$ in each sublattice [50,51]. When subjected to a current applied at an angle $\theta$ relative to the $x$-axis ($u_x = u\cos\theta,\ u_y = u\sin\theta,$ as illustrated in **Fig. S1**), the velocity of elliptical skyrmion under anisotropic field can be written as

$$v_x = \frac{(\alpha\beta D_{xx}D_{yy} + g^2)\, u\cos\theta - (\alpha-\beta)\, gD_{yy} u\sin\theta}{\alpha^2 D_{xx}D_{yy} + g^2}$$

$$v_y = \frac{(\alpha\beta D_{xx}D_{yy} + g^2)\, u\sin\theta + (\alpha-\beta)\, gD_{xx} u\cos\theta}{\alpha^2 D_{xx}D_{yy} + g^2} \tag{2}$$

Through [$C_2$‖$C_{4z}$**t**] symmetry in ATM lattice, $D_{xx}^{A} = D_{yy}^{B} = D_1$ and $D_{xx}^{B} = D_{yy}^{A} = D_2$. According to the above analysis, the transverse velocity of ATM skyrmion can be expressed as (detailed in **Note 2** of Supplemental Material [52]):

$$v_\perp = -\frac{(\alpha\text{-}\beta)\,(D_1\text{-}D_2)\, gu\cos 2\theta}{\alpha^2 D_1 D_2 + g^2} \tag{3}$$

This indicates that the SkHE can be preserved in antiferromagnetically coupled systems with zero net magnetization. The corresponding skyrmion Hall angle of the ATM skyrmion is given by:

$$\tan\phi_{SKH} = \frac{v_\perp}{v_{//}} = -\frac{(\alpha\text{-}\beta)\,(D_1\text{-}D_2)\, g\cos 2\theta}{2(\alpha\beta D_1 D_2 + g^2) + (\alpha\text{-}\beta)\,(D_1\text{-}D_2)\, g\sin 2\theta} \tag{4}$$

Evidently, the skyrmion Hall angle $\phi_{SKH}$ of ATM skyrmion is a function of the current injection angle $\theta$, which suggests the realization of A-SkHE in 2D ATM lattice. As illustrated in **Fig. 1b**, when the injected current is along the $x$- or $y$-axis, the transverse velocity of the ATM skyrmion are equal in magnitude but reverses in direction. Intriguingly, when the current is applied at specific angle $\theta = 45°$, the ATM skyrmion exhibits no transverse velocity with $\phi_{SKH} = 0$, indicating the disappearance of SkHE. We wish to stress that this mechanism for realizing A-SkHE is also applicable for 2D ATM lattice with other symmetries.

In the following, we illustrate this mechanism in Janus monolayer $V_2SeTeO$, which is a typical ATM system [63,64]. **Fig. 2a** presents the crystal structure of monolayer $V_2SeTeO$. It exhibits the space group *P*4mm. Each cell contains two V atoms. Each V atom is coordinated by two Se, two Te and two O atoms, which forms a distorted octahedral structure. The lattice constants of monolayer $V_2SeTeO$ are optimized to be $a = b = 3.94$ Å. Monolayer $V_2SeTeO$ preserves in-plane rotational ($C_{4z}$) symmetry, while its out-of-plane symmetry is broken. The rotational $C_{4z}\mathbf{t}$ symmetry establishes a connection between two nearest neighboring V atoms with anisotropic coordination environments, as shown in **Fig. 2b**, thereby satisfying the spatial symmetry conditions of altermagnetism. The phonon spectra shown in **Fig. S2a,** indicates that it is dynamically stable. Its thermal stability is further confirmed by performing ab initio molecular dynamics (AIMD) simulations, as shown in **Fig. S2b**. These results are consistent with previous studies [63,64].

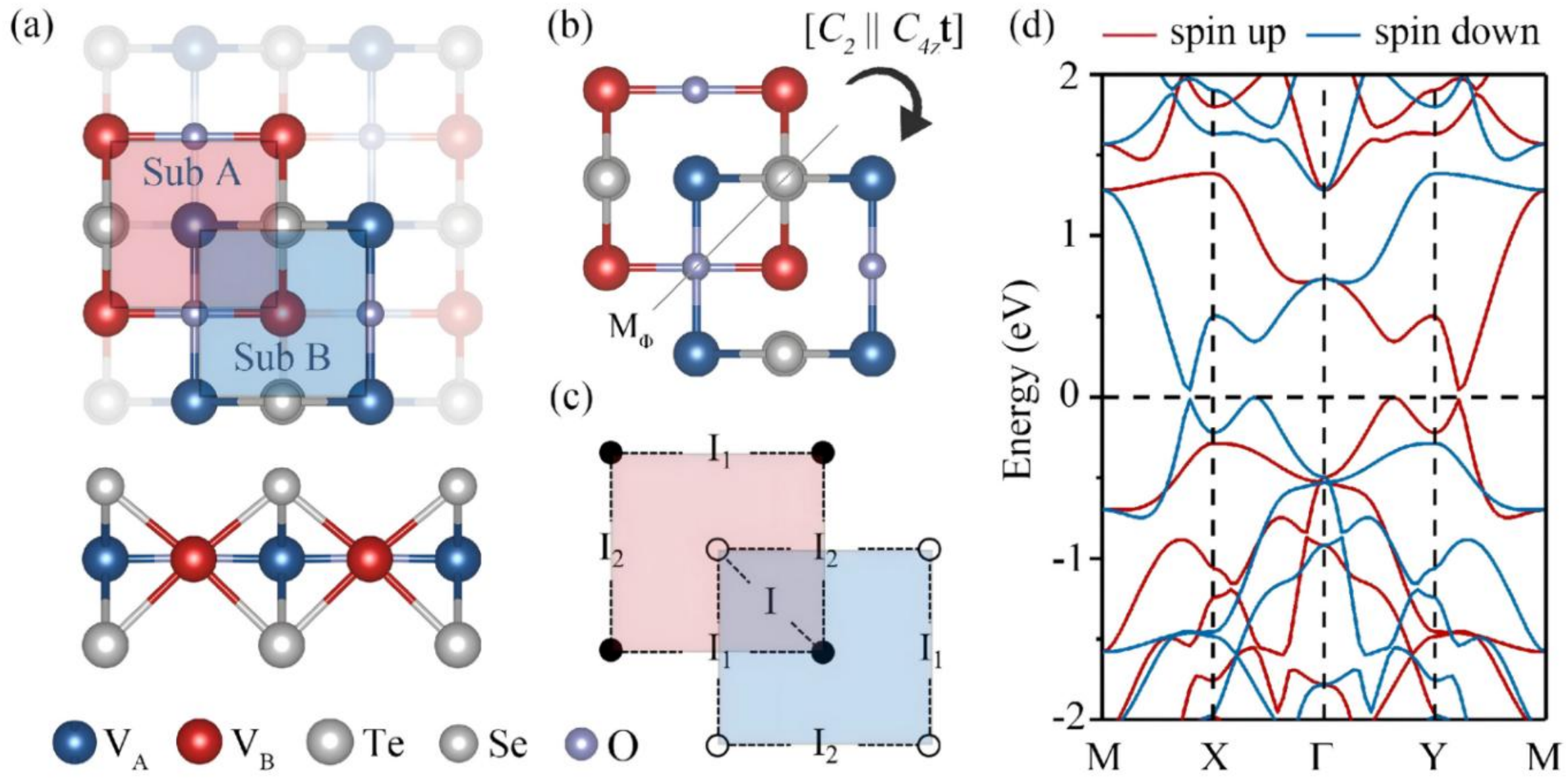

**Fig. 2** (a) Top and side views of crystal structure of monolayer $V_2SeTeO$. Red and blue tetragons in (a) indicate sublattice A and B, respectively. (b) Schematic diagrams of symmetry operation connecting two sublattices. (c) Schematic diagrams of interactions ($I_1 \neq I_2 \neq I$) within monolayer $V_2SeTeO$. (d) Band structures of monolayer $V_2SeTeO$.

The valence electronic configuration of V atom is $3d^3 4s^2$. In monolayer $V_2SeTeO$, each V atom donates three electrons to the coordinated O, Se and Te atoms, resulting in an ionic configuration of $3d^2 4s^0$. Within distorted octahedral coordination environment, Cr-3d orbitals roughly split into two groups: high-lying $e_g(d_{xy}, d_{x^2-y^2})$ and low-lying $t_{2g}(d_{xz}, d_{yz}, d_{z^2})$. The two remaining electrons of V atom half-occupy $t_{2g}$ orbitals, which results in a magnetic moment of 2 $\mu_B$. To determine the magnetic ground state, both spin-parallel and spin-antiparallel configurations of monolayer $V_2SeTeO$ are considered. The AFM coupling is found to be energetically preferred. This AFM ordering exhibits $C_2$ symmetry in spin space with the axis perpendicular to the spins, satisfying the spin space symmetry of altermagnetism. Accordingly, the magnetic lattice of monolayer $V_2SeTeO$ can be divided into two sublattices (A and B), interconnected via the $[C_2 \| C_{4z}\boldsymbol{t}]$ symmetry operation.

The band structure of monolayer $V_2SeTeO$ is shown in **Fig. 2d**. Despite possessing vanishing net magnetic moment, the system exhibits distinct spin-polarized band splitting — a hallmark characteristic of altermagnetism [65-67]. Specifically, the bands along M-X-Γ and Γ-Y-M high-symmetry paths display complete degeneracy in energy dispersion, but originate from opposite spin channels. This phenomenon is protected by the diagonal mirror ($M_\Phi$) symmetry, as shown in **Fig. 2b**, which enforces spin-momentum locking while simultaneously breaking of the combined space inversion, time reversal, and translation symmetries [65-68].

To further investigate the magnetic properties of monolayer $V_2SeTeO$, we construct a 2D atomic spin Hamiltonian model [69-71]

$$H = -\sum_{\langle i,j \rangle} J\,\boldsymbol{S}_i \cdot \boldsymbol{S}_j - \sum_n \sum_{\langle i,j \rangle_n} J_n\,\boldsymbol{S}_i \cdot \boldsymbol{S}_j - \sum_{\langle i,j \rangle} \boldsymbol{d} \cdot (\boldsymbol{S}_i \times \boldsymbol{S}_j) - \sum_n \sum_{\langle i,j \rangle_n} \boldsymbol{d}_n \cdot (\boldsymbol{S}_i \times \boldsymbol{S}_j) - K \sum_i (S_i^z)^2 \tag{4}$$

Here, $\boldsymbol{S}_{i(j)}$ is the unit vector taken from the actual atomic moment $\boldsymbol{m}_{i(j)}$ and given by $S_{i(j)} = \boldsymbol{m}_{i(j)}/|\boldsymbol{m}_{i(j)}|$. The summations $\langle i,j \rangle$ and $\langle i,j \rangle_n$ run over all inter-sublattice V atomic pairs and intra-sublattice V atomic pairs, respectively. $J$ and $J_n$ ($n$ = 1, 2) represents the inter-sublattice and intra-sublattice Heisenberg exchange coupling, respectively. $\boldsymbol{d}$ and $\boldsymbol{d}_n$ ($n$ = 1, 2) indicates the in-plane Dzyaloshinskii-Moriya interaction (DMI) between the inter-sublattice and intra-sublattice sites, respectively. The parameter $n$ is introduced to distinguish the different intra-sublattice interactions arising from the sublattice-specific anisotropic fields as illustrated in **Fig. 2c**. Specifically, $n$ = 1(2) specifies the interaction of V atomic pairs connected by O (S/Se) atoms. The last term describes the single-ion anisotropy (SIA).

As detailed in **Note 4** of Supplemental Material [52], four distinct magnetic configurations are constructed to determine the Heisenberg exchange interactions in monolayer $V_2SeTeO$. Our calculations yield exchange parameters of $J$ = -40.31 meV, $J_1$ = 31.10 meV and $J_2$ = 28.32 meV. These results reveal AFM coupling for inter-sublattice interactions, while intra-sublattice coupling exhibits anisotropic FM behavior. Regarding DMI, only in-plane components $\boldsymbol{d}$ and $\boldsymbol{d}_n$ are considered, as out-of-plane component does not contribute to the spiral magnetic arrangement. Following the Moriya's symmetry rules [72], in-plane DMI vectors of monolayer $V_2SeTeO$ are oriented perpendicular to their respective bonding directions. These are calculated to be $\boldsymbol{d}$ = 0.81 meV, $\boldsymbol{d}_1$ = -0.56 meV and $\boldsymbol{d}_2$ = -0.13 meV, where the positive values denote counterclockwise direction. Clearly, the inter-sublattice DMI adopts a counterclockwise orientation, whereas the intra-sublattice DMI exhibits a clockwise direction (**Fig. S4**). While for SIA, the parameter $K$ is determined to be 0.03 meV, indicating the preference for out-of-plane magnetization alignment.

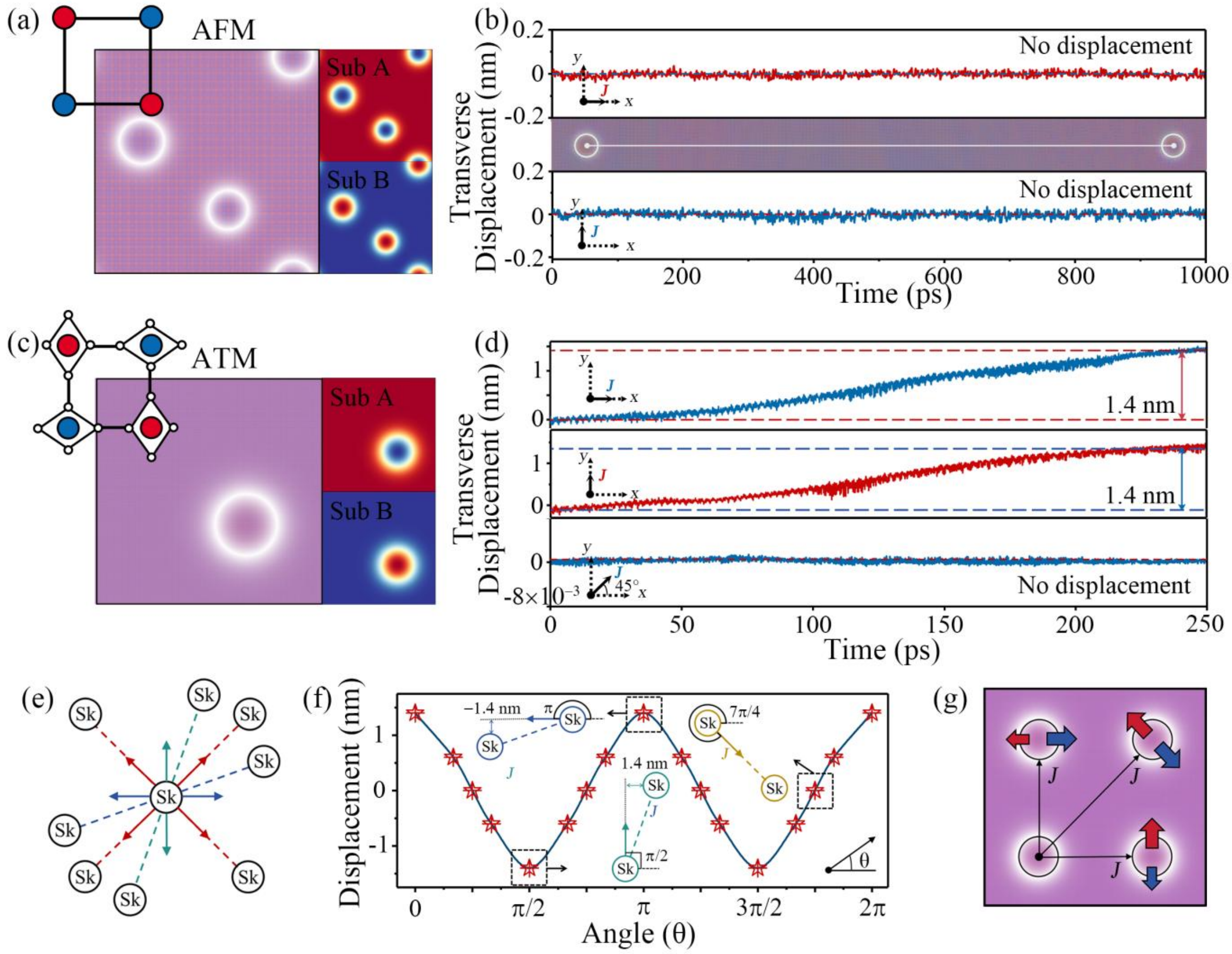

**Fig. 3** Spin textures of monolayer $V_2SeTeO$ (a) without and (c) with considering the effective sublattice-specific anisotropic fields. Time evolution of the transverse displacement of skyrmion under current applied in different directions for monolayer $V_2SeTeO$ (b) without and (d) with considering the effective sublattice-specific anisotropic field. Middle panels in (b) are the corresponding illustrations of real-space trajectories for skyrmions. (e) Schematic diagram of motion directions for ATM skyrmions under currents applied in different directions. (f) The relationship between ATM skyrmion

transverse displacement (at 1000 ps) and current injection angle $\theta$. The colored arrows illustrate directions of current injection and corresponding colored dash lines illustrate the directions of skyrmion motion in (e,f). (g) Schematic diagram of A-SkHE in altermagnets, wherein red (blue) arrows represent the transverse velocity of sublattice A (B).

Based on the first-principles parameterized Hamiltonian, we perform atomic spin model simulations to explore the spin textures in monolayer $V_2SeTeO$. This system can be conceptualized as an AFM framework with additional sublattice-specific anisotropic fields. To isolate the unique characteristics of altermagnetism arising from these anisotropic fields, we first carry out control simulations by intentionally suppressing the anisotropic contributions, thereby reducing the system to a conventional AFM state. As shown in **Fig. 3a**, the resulting spin texture features AFM skyrmion structure comprising two nested FM skyrmions with antiparallel spin orientations. Crucially, the coexistence of $[C_2\|\mathbf{t}]$ and $[C_2\|\mathbf{P}]$ symmetries between the two sublattices enforces a mutual cancellation of transverse velocity components for the constituent FM skyrmions. This symmetry-protected mechanism suppresses SkHE, as predicted by our theoretical framework. To quantitatively validate this behavior, we implement spin dynamics simulations via **Eq. (1)**, tracking the time evolution of the transverse displacement of skyrmion in presence of an applied current. As shown in **Fig. 3b**, regardless of whether the current is applied in the x-direction or the y-direction, the central position of AFM skyrmion oscillates around its equilibrium position with no observed transverse deviation. The real-space trajectory diagram (middle panel in **Fig. 3b**) further illustrates this confined oscillatory motion, confirming the absence of SkHE-driven transverse drift.

**Fig. 3c** displays the spin textures of monolayer $V_2SeTeO$, with taking effective sublattice-specific anisotropic fields into account. Notably, this configuration reveals the emergence of a characteristic ATM skyrmion—a topological quasiparticle distinct from conventional AFM skyrmion. As compared with the AFM skyrmion shown in **Fig. 3a**, under ATM ordering, the skyrmion density is decreased and the skyrmion size is enlarged. Current-driven dynamics further differentiate this state. Under $+x$ axis current injection, the ATM skyrmion develops a pronounced transverse velocity component, achieving a ~1.4 nm displacement along $+y$ direction at 250 ps (**Fig. 3d**). When the current is rotated to the +y axis, this Hall-like motion exhibits directionality inversion, with the skyrmion migrating 1.4 nm along the +x direction under equivalent timescales.

The detailed real-space trajectories of the ATM skyrmion motion are quantitatively analyzed in **Fig. S5 a** and **c**. It can be seen that the skyrmion Hall angle maintains identical absolute values but undergoes sign inversion when switching current direction between $+x$ and $+y$ axis. This directional switching phenomenon originates from the sublattice-specific anisotropic fields of monolayer $V_2SeTeO$. Specifically, as illustrated in **Fig. 3e**, when the current is applied along the *x*-direction, the skyrmion in sublattice A exhibits a greater transverse velocity, whereas for a current applied along the *y*-direction, the skyrmion in sublattice B exhibits a larger transverse velocity. The result is in excellent agreement with the above theoretical analysis. During our simulations, we explicitly monitor the shape of skyrmion throughout its current-driven motion as shown in **Fig. S5**. It confirms that the shape of skyrmions remains remarkably stable with no distortion, stretching, or breathing modes under the applied current densities.

To provide a systematically characterization of the motion of ATM skyrmions under currents applied in different directions, we perform spin dynamics simulations mapping the relationship between the skyrmion transverse displacement and the angle of applied current. As displayed in **Fig. 3f**, the obtained results precisely match the functional form predicted by **Eq. (3)**. To provide a more intuitive representation of the simulation results, **Fig. 3g** presents a schematic illustration of their movement trajectories. Evidently, such current-direction-dependent transverse response starkly contrasts with the SkHE-suppressed behavior observed in the isotropic AFM phase, directly manifesting A-SkHE in 2D ATM lattice.

Another intriguing feature we can see from **Fig. 3d** and **f** is that when the current injection angle is $\theta = (2n+1)\,\pi/4 \;\; (n \in \mathrm{N})$, the ATM skyrmion exhibits no transverse displacement, leading to complete suppression of SkHE (**Fig. S5b**). This compensation effect is primarily attributed to the fact that, at this specific angle, the transverse velocities of skyrmions in the two sublattices are equal in magnitude but opposite in direction, as illustrated in **Fig. 3e**.

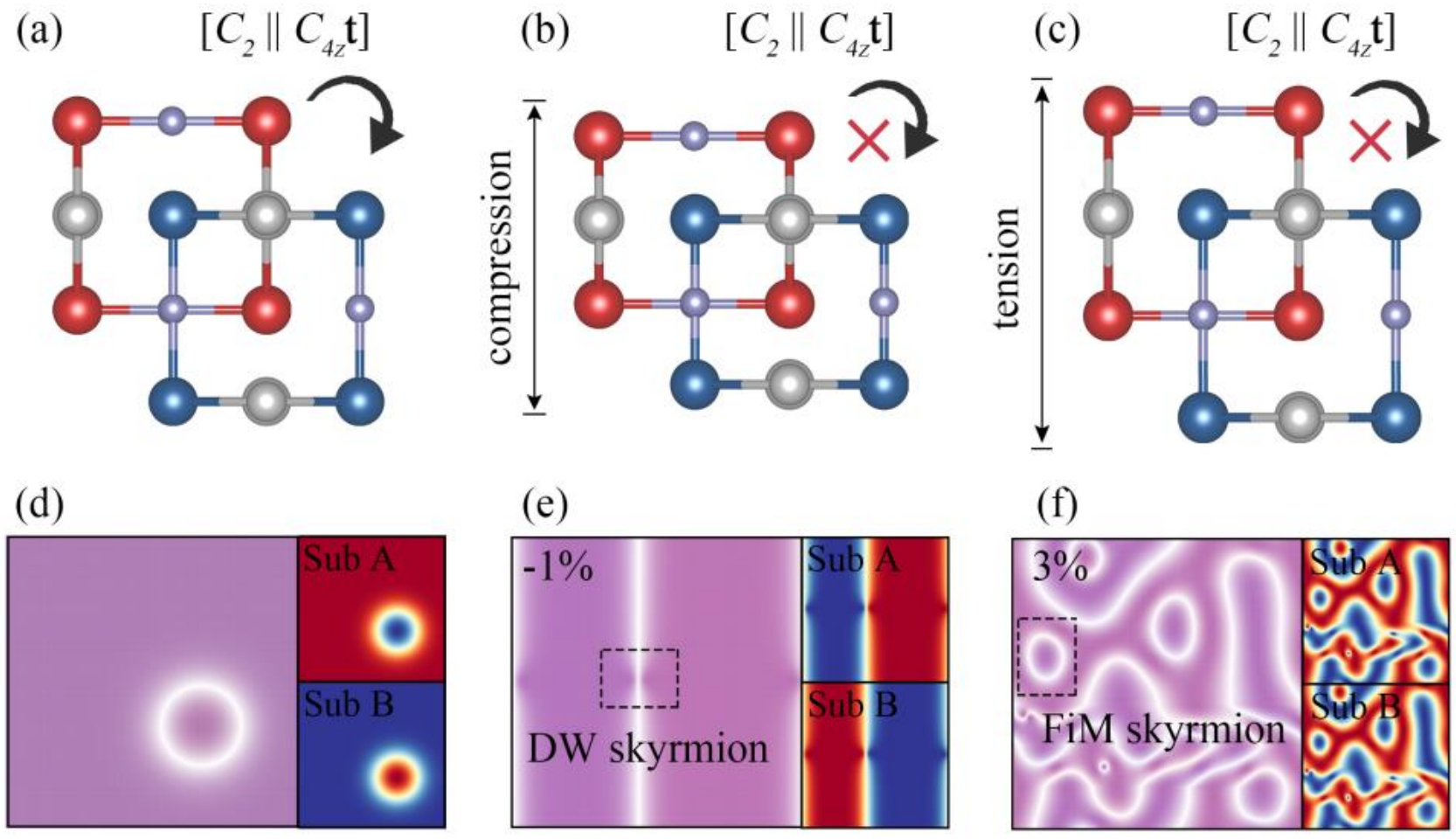

**Fig. 4** Crystal structures and symmetries of monolayer $V_2SeTeO$ (a) without strain, with (b) compression uniaxial strain and (c) tensile uniaxial strain. The corresponding real-space spin textures of monolayer $V_2SeTeO$ (d) without strain, with (e) compression uniaxial strain and (f) tensile uniaxial

strain.

**Table 1.** The DMI component $d_{\parallel}$ along the direction connecting two atoms, DMI component $d_{\perp}$ perpendicular to the direction connecting two atoms, Heisenberg exchange interaction $J$ and SIA parameter $K$ for monolayer $V_2SeTeO$ under different strain.

| | $d_{\parallel}$ /meV | $d_{\perp}$ /meV | $J$ /meV | $K$ /meV |
|---|---|---|---|---|
| -3% | <0.001 | 0.33 | 40.78 | -0.15 |
| -1% | <0.001 | 0.49 | 40.71 | 0.03 |
| 1% | <0.001 | 0.71 | 38.62 | 0.07 |
| 3% | 0.78 | 1.04 | 34.51 | -0.01 |

We wish to stress that the emergence of ATM skyrmions and resultant A-SkHE in monolayer $V_2SeTeO$ is inherently protected by $[C_2 \| C_{4z}\mathbf{t}]$ symmetry. Upon breaking the $[C_2 \| C_{4z}\mathbf{t}]$ symmetry, the altermagnetism of monolayer $V_2SeTeO$ will undergo a transition into ferrimagnetism, thereby enabling a skyrmion phase transition. To verify it, as depicted in **Fig. 4b** and **c**, uniaxial compressive and tensile strains are applied to monolayer $V_2SeTeO$. According to above results, we focus on the inter-sublattice interactions for the strained cases for simplicity.

Since uniaxial strain breaks the mirror symmetry $M_{\Phi}$, the DMI of monolayer $V_2SeTeO$ is no longer constrained within the perpendicular bisector plane of the atomic bond according to Moriya's symmetry rules [72]. Consequently, the system exhibits two distinct components: $d_{\parallel}$ along the atomic bond direction and $d_{\perp}$ perpendicular to it. The calculated magnetic parameters are summarized in **Table 1**. Through the atomic spin model simulations, domain-wall (DW) skyrmion emerges in monolayer $V_2SeTeO$ under -1% compression strain (**Fig. 4e**). DW skyrmions are confined within DW while exhibiting a well-defined topological charge, ensuring their topological protection. Therefore, a phase transition between ATM skyrmions and DW skyrmions is achieved by applying compressive strain. On the other hand, when a 3% tensile strain is applied, the monolayer $V_2SeTeO$ exhibits a mixed state comprising labyrinth domains and ferrimagnetic (FiM) skyrmions as shown in **Fig. 4f**. This complex phase coexistence suggests that the switch between ATM skyrmion and FiM skyrmion can also be achieved by the application of tensile strain on monolayer $V_2SeTeO$.

## Conclusion

To summarize, our investigation establishes A-SkHE in 2D AFM lattice by exploring ATM skyrmions.

With the help of symmetry analysis and theoretical modeling, we uncover the mechanism of A-SkHE in 2D ATM systems and a quantitative correlation between the transverse velocity of ATM skyrmions and the orientation of the applied current. First-principles calculations and micromagnetic simulations further validate this mechanism in monolayer $V_2SeTeO$. These findings provide critical insights into the exploration of topological spintronics in 2D altermagnets.

## Data Availability Statement

The data that support the findings of this article are included within the article and its Supplemental Material.

## Acknowledgments

This work is supported by the National Natural Science Foundation of China (No. 12274261) and Taishan Young Scholar Program of Shandong Province.